# Polynomial time and space quantum algorithm for the simulation of non-Markovian quantum dynamics

Avin Seneviratne[a], Peter L. Walters[b], Fei Wang[b,c*]

[a]Department of Physics and Astronomy, George Mason University, 4400 University Drive, Fairfax, Virginia 22030, USA

[b]Department of Chemistry and Biochemistry, George Mason University, 4400 University Drive, Fairfax, Virginia 22030, USA

[c]Quantum Science and Engineering Center, George Mason University, 4400 University Drive, Fairfax, Virginia 22030, USA

These two autors (a,b) contribute equally to the work.
*Email: fwang22@gmu.edu

## Abstract

In this work, we developed an efficient quantum algorithm for the simulation of non-Markovian quantum dynamics, based on the Feynman's path integral formulation. The algorithm scales polynomially with the number of native gates and the number of qubits, and has no classical overhead. It demonstrates the quantum advantage by overcoming the exponential cost on classical computers. In addition, the algorithm is efficient regardless of whether the temporal entanglement due to non-Markovianity is low or high, making it a unified framework for simulating non-Markovian dynamics in open quantum system.

## I.      Introduction

Quantum computing has demonstrated its supremacy in solving classically hard problems, with the culmination of Shor's factoring[1,2] and Grover's search.[3] Considering the current stage of the technological development, quantum simulation appears to be the first immediate application of quantum computing.[4,5] Numerous efforts have been devoted to developing quantum algorithms for simulating closed quantum systems,[6–11] whereas less work has focused on open quantum systems. However, open quantum system dynamics have important ramifications for the charge and excitation energy transfer processes in the condensed phase environment.[12–19] The non-Markovian regime simulation has been particularly challenging on classical computers, often resulting in exponential scaling with respect to the memory length and system size. The reason is that with the system keeping track of its past, the number of possible states grow exponentially with the propagation time. The advantage of quantum computers is that the exponentially large Hilbert space can be encoded in a linear number of qubits. In recent years, several groups have devoted effort in developing quantum algorithms for non-Markovian quantum dynamics. Head-Marsden et al.[20] used ensemble of Lindblad trajectories originating from different times to capture the non-Markovian behavior. The Lindblad trajectories are simulated by the quantum computer. The number of trajectories can be large for high non-Markovian regime. Wang et al.[21] used the Nakajima-Zwanzig equation to generate the non-Markovian superoperator and implemented the unitary version of the superoperator on the quantum computer. The major computational cost is the construction of the superoperator on the classical computer by solvoing the integrated differential equation. Seneviratne et al.[22] used the Feynman-Vernon's influence functional to construct the Kraus operators associated with the superoperator. The advantage is that the low dimension of the Kraus operator is more NISQ-friendly whereas the main cost is the computation of the singular value decomposition on the classical computer. Walters et al.[23] recently have implemented the Feynman path integral scheme on the quantum computer, with the major cost of calculating the eigenvalues of the propagator. In addition, Tsakanikas *et al.*[24] have developed a variatonal quantum algorithm that captures the non-Markovian dynamics through the ensemble averaged classical path. The main cost is solving the differential equation necessary for the parameter update and the Monte Carlo sampling on the classical computer. In this work, we improve upon the previous path integral method by competely staying away from the classical overhead. We will show that the quantum algorithm we developed is both polynomial in time and spce, circumventing the exponential scaling on classical computers. In addition, the algorithm is equally efficient outside of the low entanglement regime where tensor-network method[25,26] can apply. The paper is organized as the following. In Section II, we review the path integral approach for non-Markovian quantum dynamics. In Section III, we delineate the quantum algorithm we developed. In section IV, we show our simulation results and offer discussions. In Section V, we make some conclusion remarks.

## II. Path integral formulation

In this section, we briefly review the Feynman's path integral formulation for non-Markovian quantum dynamics, and we will focus on the model of a quantum system linearly coupled to its harmonic bath. The reduced density matrix (RDM) is commonly used to describe the dynamics of open quantum systems. The Hamiltonian of a $m$-state system linearly coupled to its harmonic bath can be described as the following:

$$H_s = \sum_{s=1}^{m} s|s\rangle\langle s| + \sum_{s,s'} V_{s,s'} |s\rangle\langle s'| \tag{2.1}$$

$$H_b = \sum_j \hbar \omega_j a_j^\dagger a_j \tag{2.2}$$

$$H_{s-b} = -\sum_{s=1}^{m} s|s\rangle\langle s| \left[ \sum_j c_j \sqrt{\frac{\hbar}{2m_j\omega_j}} (a_j^\dagger + a_j) \right] \tag{2.3}$$

where $H_s$ and $H_b$ denote the system and bath Hamiltonian, respectively, and $H_{s-b}$ the system bath coupling. The system states $|s\rangle$ are usually expressed in the diabatic basis that diagonalizes the position operator. This representation naturally fits into the position representation of the path integral formulation. The coupling strength $c_j$ and the frequency $\omega_j$ collectively define the spectral density[27,28]:

$$J(\omega) = \frac{\pi}{2} \sum_j \frac{c_j^2}{m_j\omega_j} \delta(\omega - \omega_j) \tag{2.4}$$

Drude and Ohmic spectral density are the routinely used model spectra, which have a characteristic peak and a cutoff. For real molecular systems, the spectral density can be obtained from the Huang-Rhys factors[29] and molecular dynamics simulations[12,30].

The RDM in Feynman's path integral formulation is expressed as the system's bare propagator multiplied by the influence functional[31]:

$$\rho_{red}(s_N^+, s_N^-; t) = \int ds_0^+ \int ds_1^+ \cdots \int ds_{N-1}^+ \int ds_0^- \int ds_1^- \cdots \int ds_{N-1}^-$$

$$\times \langle s_N^+|e^{-iH_s\Delta t/\hbar}|s_{N-1}^+\rangle \cdots \langle s_1^+|e^{-iH_s\Delta t/\hbar}|s_0^+\rangle \langle s_0^+|\rho_s(0)|s_0^-\rangle$$

$$\times \langle s_0^-|e^{iH_s\Delta t/\hbar}|s_1^-\rangle \cdots \langle s_{N-1}^-|e^{iH_s\Delta t/\hbar}|s_N^-\rangle$$

$$\times I(s_0^+, s_1^+, \cdots, s_{N-1}^+, s_N^+, s_0^-, s_1^-, \cdots, s_{N-1}^-, s_N^-; \Delta t) \quad (2.5)$$

The $\{s_0^+, s_1^+, \cdots, s_N^+\}$ and $\{s_0^-, s_1^-, \cdots, s_N^-\}$ are the discretized forward and backward paths, and $\langle s_0^+|\rho_0(0)|s_0^-\rangle$ is the system's initial condition, with $\Delta t$ the Trotter time step.

The influence functional has the expression

$$I = \exp\left[-\frac{1}{\hbar}\sum_{k=0}^{N}\sum_{k'=0}^{k}(s_k^+ - s_k^-)(\alpha_{kk'}s_{k'}^+ - \alpha_{kk'}^*s_{k'}^-)\right] \quad (2.6)$$

where the $\alpha_{kk'}$ coefficients (see Appendix A) are derived by Makri.[32]

In continuous time, the influence functional assumes the form of[28,33]

$$IF = \exp\left\{-\frac{1}{\hbar}\int_0^t dt' \int_0^{t'} dt'' [s^+(t') - s^-(t')][\alpha(t'-t'')s^+(t'') - \alpha^*(t'-t'')s^-(t'')]\right\} \quad (2.7)$$

where

$$\alpha(t'-t'') = \frac{1}{\pi}\int_0^\infty d\omega J(\omega)\left[\coth\left(\frac{\hbar\omega\beta}{2}\right)\cos(\omega t' - \omega t'') - i\sin(\omega t' - \omega t'')\right] \quad (2.8)$$

This time correlation function $\alpha(t'-t'')$ is non-local and is root of the non-Markovian character of the dynamics. For the condensed phase environment where the bath is composed of a broad range of frequencies, $\alpha(t'-t'')$ has a finite time span.[32]

Although modeling the environment's degrees of freedom using harmonic oscillators is the major assumption, central limit theorem[34] guarantees that this Gaussian response can be widely applied to many condensed phase systems[12–19,35–40].

### III. Quantum algorithm
### A. Tensor product structure

As a preliminary, it is convenient to re-express equation (2.5) in a tensor product form. Defining a short time unitary propagator as

$$K(s_{k+1}^\pm, s_k^\pm) = \langle s_{k+1}^+|e^{-\frac{iH_s\Delta t}{\hbar}}|s_k^+\rangle\langle s_k^-|e^{\frac{iH_s\Delta t}{\hbar}}|s_{k+1}^-\rangle \quad (3.1)$$

then the system's bare propagation up to $N\Delta t$ is

$$\mathcal{U} = \prod_{k=0}^{N-1} K(s_{k+1}^\pm, s_k^\pm) \quad (3.2)$$

If the memory spans $L\Delta t$ where $L \leq N$, the influence functional can be rewritten as[32]

$$IF = \prod_{k=0}^{N} I_{kk}(s_k^{\pm}, s_k^{\pm}) \prod_{k=0}^{N-1} I_{k,k+1}(s_{k+1}^{\pm}, s_k^{\pm}) \cdots \prod_{k=0}^{N-L} I_{k,k+l}(s_{k+L}^{\pm}, s_k^{\pm}) \quad (3.3)$$

where

$$I_{k,k'}(s_{k'}^{\pm}, s_k^{\pm}) = \exp\left[-\frac{1}{\hbar}(s_{k'}^{+} - s_{k'}^{-})(\alpha_{k'k} s_k^{+} - \alpha_{k'k}^{*} s_k^{-})\right] \quad (3.4)$$

The RDM at $N\Delta t$ is therefore expressed as a tensor product of the RDM at $t = 0$, a unitary operator, and the influence functional tensor,

$$\rho_{red}(s_N^+, s_N^-) = \sum_{s_{N-1}^{\pm}} \cdots \sum_{s_0^{\pm}} \langle s_0^+ | \rho_0(0) | s_0^- \rangle \times \mathcal{U} \times IF \quad (3.5)$$

Timewise, the $I_{kk}(s_k^{\pm}, s_k^{\pm})$ describes self-interactions. The short time unitary operator $K(s_{k+1}^{\pm}, s_k^{\pm})$ and the influence functional tensor $I_{k,k+1}(s_{k+1}^{\pm}, s_k^{\pm})$ describe the nearest-time coupling. All other influence functional tensors describe non-local temporal interactions. Figure 1 shows a picture representation.

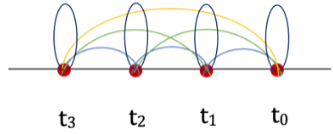

Figure 1. Non-Markovian temporal coupling.

This tensor-product structure forms the basis for the following quantum algorithm development. We can create equal superposition states that represent all the possible paths. For instance, to propagate one timestep, a two-state system (denoted by the eigenvalues, e.g. 0 and 1, of the position-like states) has the possible forward-backward paths of (00;00), (00;01), (00;10), (00;11), (01;00), etc. These superposition states can be prepared using two qubits. Since the $\{s_k^{\pm}\}$ in the path integral expression are the positions and the states just prepared represent position states, the short time unitary propagator $K(s_{k+1}^{\pm}, s_k^{\pm})$ and the influence functional tensor $I_{k,k'}(s_{k'}^{\pm}, s_k^{\pm})$ are diagonal in this basis. Therefore, the state generated by applying $\langle s_0^+ | \rho_0(0) | s_0^- \rangle \times \mathcal{U} \times IF$ to the equal superposition states $\sum |i\rangle$ is

$$\sum_i A_i |i\rangle \quad (3.6)$$

where $A_i$ is a complex number that represents the amplitude associated with the specific path $|i\rangle$. The value of $A_i$ corresponds to putting specific path-list $\{s_k^{\pm}\}$ into equation (3.5). To obtain the RDM, the sum of amplitude, $\sum_i A_i$, is performed. This can be done using the Hadamard gate. The details will be described in the Circuit Structure section (C.).

## B. Unitary dilation

On quantum computers, the operations have to be unitary. The $I$ tensor in the influence functional is a non-unitary diagonal and can be dilated to a unitary one in the following way.[41] First, define a unitary diagonal operator associated with the elements of $I$ by

$$U_\Sigma = \begin{pmatrix} I_+ & 0 \\ 0 & I_- \end{pmatrix} \tag{3.7}$$

where

$$I_{\pm j} = \sigma_j \pm i \sqrt{\frac{1 - |\sigma_j|^2}{|\sigma_j|^2}} \sigma_j \tag{3.8}$$

and $\sigma_j$ are the elements of $I$. Equation (3.7) assumes that $|\sigma_j| < 1$. If this condition is not met, the diagonal elements $\sigma_j$ can be rescaled by the largest singular value of $I$. The rescaled value can ensure the validity of (3.8).

The circuit construction for the $U_\Sigma$ is shown in Figure 2, where the ancilla qubit implements the Hadamard gates.[41]

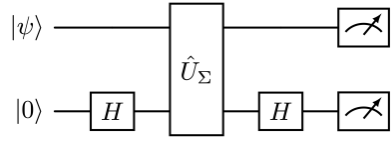

Figure 2. Circuit for unitary dilation.

The resulting states are

$$\frac{1}{2} \begin{pmatrix} (I_+ + I_-)|\psi\rangle \\ (I_+ - I_-)|\psi\rangle \end{pmatrix} = \begin{pmatrix} I|\psi\rangle \\ |\varphi\rangle \end{pmatrix} \tag{3.9}$$

in which the ancilla in the $|0\rangle$ state implements $I|\psi\rangle$, whereas the $|1\rangle$ state measurement (labeled $|\varphi\rangle$) is discarded. The measurement results show the statistics of $I|\psi\rangle\langle\psi|I^\dagger$ in the computational basis.

The diagonal unitary, equation (3.8), can be implemented efficiently on a quantum computer using the Walsh series representation.[42] Below we summarize briefly the key construction. First, express the integer $j, k \in [0, 2^{q-1})$ in its binary and dyadic expansion, respectively,

$$j = \sum_{i=1}^{q} j_i 2^{i-1} \tag{3.10}$$

$$k = \sum_{i=1}^{q} k_i 2^{n-i} \tag{3.11}$$

where $q$ is the number of qubits required to implement equation (3.7).

Define a matrix $W_{jk}$ and a vector $f_k$ as

$$W_{jk} = (-1)^{\sum_{i=1}^{q} j_i k_i} \tag{3.12}$$

$$f_k = (\ln U_\Sigma)_k \tag{3.13}$$

The Walsh coefficient $a_j$ is obtained from the Walsh-Fourier transform of $f_k$,

$$a_j = \frac{1}{2^q} \sum_{k=1}^{2^q} W_{jk} f_k \tag{3.14}$$

Then the unitary diagonal $U_\Sigma$ can be expressed as

$$U_\Sigma = \prod_{j=1}^{2^q} e^{i a_j \hat{Q}_j} \tag{3.15}$$

with the Walsh operator $\hat{Q}_j$ given as the tensor product of Pauli Z gates,

$$\hat{Q}_j = (\sigma_{z_1})^{j_1} \otimes (\sigma_{z_2})^{j_2} \otimes \cdots \otimes (\sigma_{z_q})^{j_q} \tag{3.16}$$

where $j_i$ is in equation (3.10).

The circuit for the exponentiation of tensor product of Pauli gates can be constructed[43] and further optimized[42] with the Gray code ordering to minimize the number of CNOT gates. The details of the circuit construction are given in Appendix B.

### C. Circuit structure

In this section, we present two algorithms. We will use a two-level system as an example, and the method can be easily extended to a multi-level system.

**(a) Algorithm I**

In the plain version of equation (3.5), each term as in (3.1) and (3.4) can be separately implemented on a quantum machine with the proper dilation. Figure 3 shows the circuit for the propagation up to $2\Delta t$. In the Figure, the 0, 1 and 2 in the qubit indicate the time point. The letter f and b represent "forward" and "backward" states. If the intial state is a pure state and population dynamics is probed, then the starting time-point ($t = 0$) and the final time-point ($t = 2\Delta t$) will simply be the computational basis. For instance, if initial state is $\langle 0|\rho(0)|0\rangle = 1$ and the final state we intend to probe is $\langle 0|\rho(2\Delta t)|0\rangle$, then the two end-point states would be $|0_f\rangle|0_b\rangle = |0\rangle|0\rangle$ and $|2_f\rangle|2_b\rangle = |0\rangle|0\rangle$. For a two-state system, the other state's population is obtained by $\langle 1|\rho(2\Delta t)|1\rangle$ with the final state preparation $|2_f\rangle|2_b\rangle = |1\rangle|1\rangle$. As a consequence, only the intermediate time points need all the superposition states which can be achieved by the Hadamard gate, as shown in Figure 3 for qubit $|1_f\rangle|1_b\rangle$. In this Figure, the $U$ and $U^\dagger$ correspond to the the forward and backward unitary propagator as in equation (3.1). However, because they are now reshaped to have a diagonal form, they lose their unitarity and therefore require dilation. The ancila qubit $|A_1\rangle$ and $|A_2\rangle$ serve this purpose and the Hadamard gates at the two ends implement the circuit of Figure 2. If the qubit line is stoped by the long rectangular block (i.e., the unitary operator), it means the operator is acting on that qubit; if the qubit line goes through the long rectangular block, the operator does not act on that qubit. As an example, the first $U$ operator only operates on qubit $|0_f\rangle$, $|1_f\rangle$ and $|A_1\rangle$, whereas the first $U^\dagger$ only operators on $|0_b\rangle$, $|1_b\rangle$ and $|A_2\rangle$. The $I_{k,k'}$ operators implement the influence functional tensor. As it turns out, the bare propagators and the self-interaction operators $I_{kk}$ are two qubit operators (plus one ancila qubit for dilation) and the other $I_{k,k'}$ operators are four qubit operators (plus one ancila). Therefore, all the operators are local and are expected to have shallow circuit structure. At this point, the outcome is equation (3.6), each basis carrying a complex value associated with a specific path. The essenece of the path integral (i.e., path sum) is that $\sum_i A_i$ gives the value of the final time-step RDM. Therefore, the last step of the algorithm is to aggregate all the complex values $A_i$ into one quantum state, and this is done by the Hadamard gates, as is shown in Figure 3 at the end of qubit $|1_f\rangle$ and $|1_b\rangle$. The resulting state is $(\sum_i A_i |00 \cdots 00\rangle)$ + other orthogonal states. Combining with the result of equation (3.9), the measurement statistics of all $|0\rangle$ states is proportional to $|\langle 0|\rho(N\Delta t)|0\rangle|^2$.

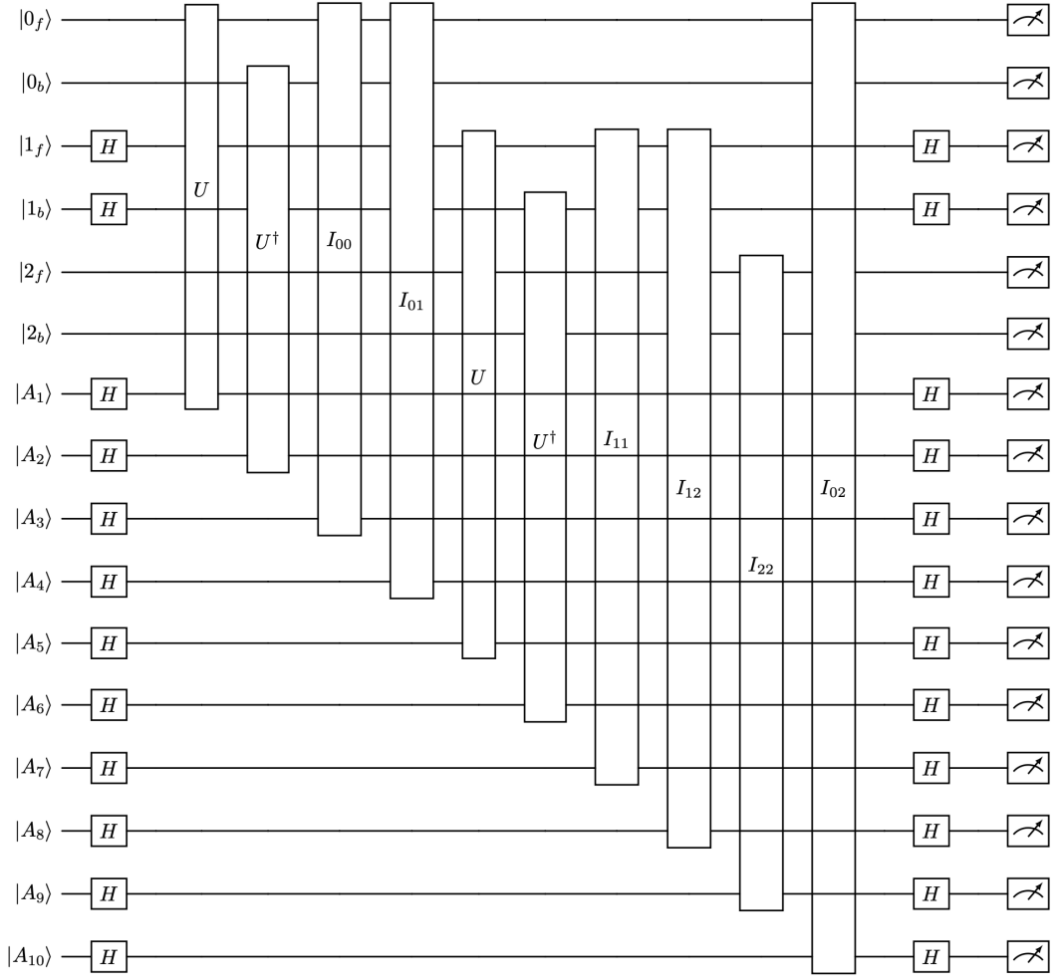

Figure 3: Circuit for 2 time steps evolution.

The way it stands, the circuit can be made more compact by combining the two-qubit operators with the four qubit operators, as long as they operate on the same qubits. For instance, for the propagation up to $2\Delta t$, we can define the combined operator $I^C$ as

$$I^C_{01} = U \times U^\dagger \times I_{00} \times I_{01} \tag{3.17}$$

$$I^C_{12} = U \times U^\dagger \times I_{11} \times I_{12} \tag{3.18}$$

$$I^C_{02} = I_{02} \times I_{22} \tag{3.19}$$

Here for conciseness, we are not introducing new symbols, but be mindful that two qubit terms in (3.17) – (3.19) such as $U$ and $I_{00}$ should tensor-multiply the identity matrix to make the dimension the same as the four qubit term such as $I_{01}$. With this process of compaction, the two qubit circuits are dropped and only the four qubit operators remain, as is shown in Figure 4. In addition, the number of ancila qubits are also reduced.

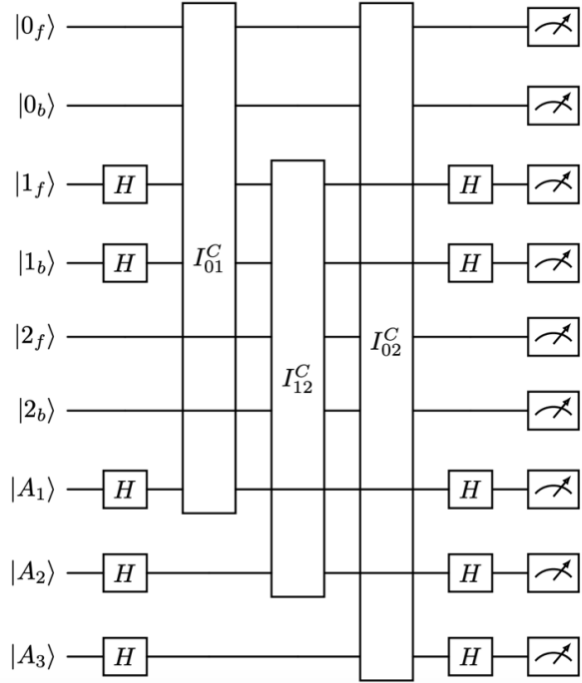

Figure 4. Compact circuit for 2 time steps evolution for population $|0\rangle$.

One subtlety is that since rescaling might be needed for the dilation (explained at the beginning of Section B), the measurement statistics of $|00 \cdots 00\rangle$ could be the rescaled value of $|\langle 0|\rho(N\Delta t)|0\rangle|^2$. Instead of keeping track of the rescaling factors, we can equally contruct another circuit that probe the population of $\langle 1|\rho(N\Delta t)|1\rangle$, as shown in Figure 5. The first two X gates on the $|2_f\rangle$ and $|2_b\rangle$ change the end-time state from $|0\rangle$ to $|1\rangle$, and the last two X gates conveniently guarantee that useful statatics are still contained in the measurement of all $|0\rangle$. As the $I^C$ operators are the same for both the circuits in Figure 4 and 5, the rescaling factor is the same. Therefore, the ratio of the statitics of the $|00 \cdots 00\rangle$ measurement result gives the ratio of $|\langle 0|\rho(N\Delta t)|0\rangle|^2 \div |\langle 1|\rho(N\Delta t)|1\rangle|^2$. Together with the identity $\langle 0|\rho(N\Delta t)|0\rangle + \langle 1|\rho(N\Delta t)|1\rangle = 1$, the population dynamics can be otained.

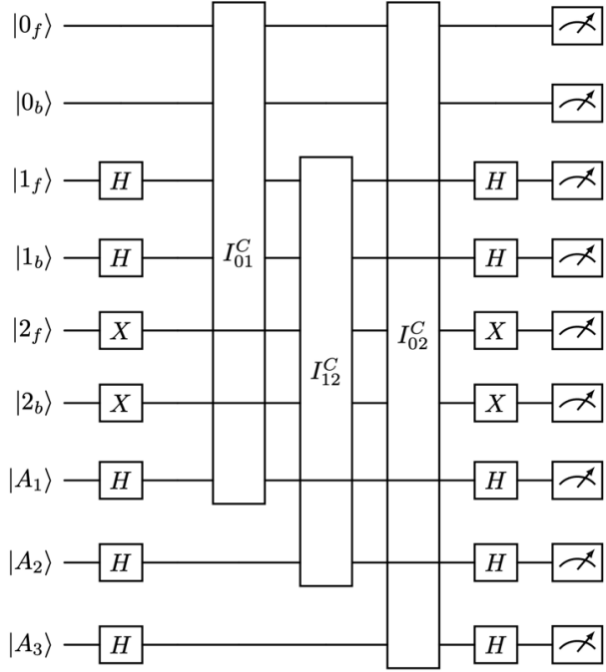

Figure 5. Compact circuit for 2 time steps evolution for population $|1\rangle$.

**(b) Algorithm II**

To alleviate the measurement overhead, we further construct the multiple-controlled-NOT gate, i.e., the Toffoli gate, to transfer the outcome of $|00\cdots 00\rangle$ into one qubit,[43] and the measurement is only performed in this ancillary qubit. We call this one-qubit measurement scheme, and the circuit structure is shown is Figure 6 and 7 for the population dynamics of $\langle 0|\rho(N\Delta t)|0\rangle$ and $\langle 1|\rho(N\Delta t)|1\rangle$.

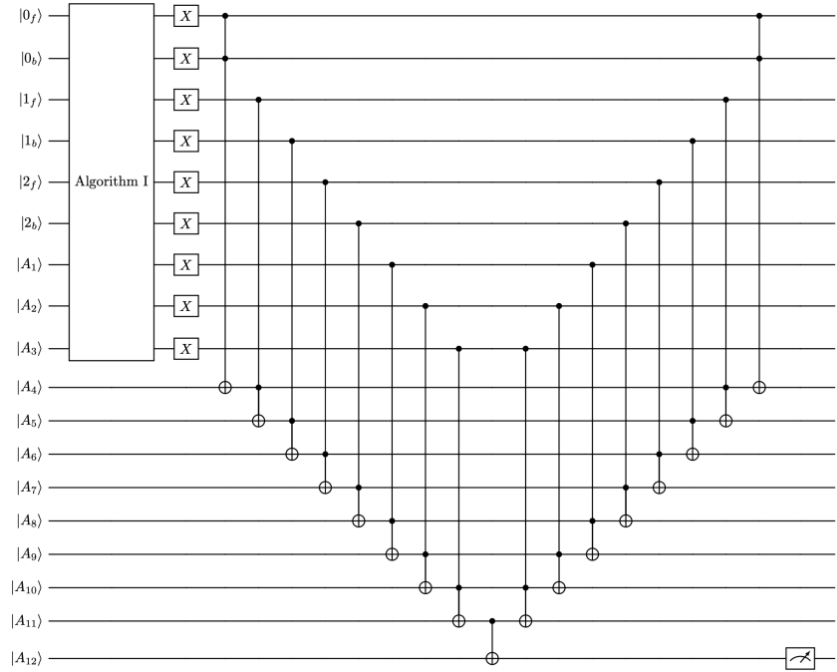

Figure 6. One-qubit measurement scheme for 2 time steps evolution for population $|0\rangle$.

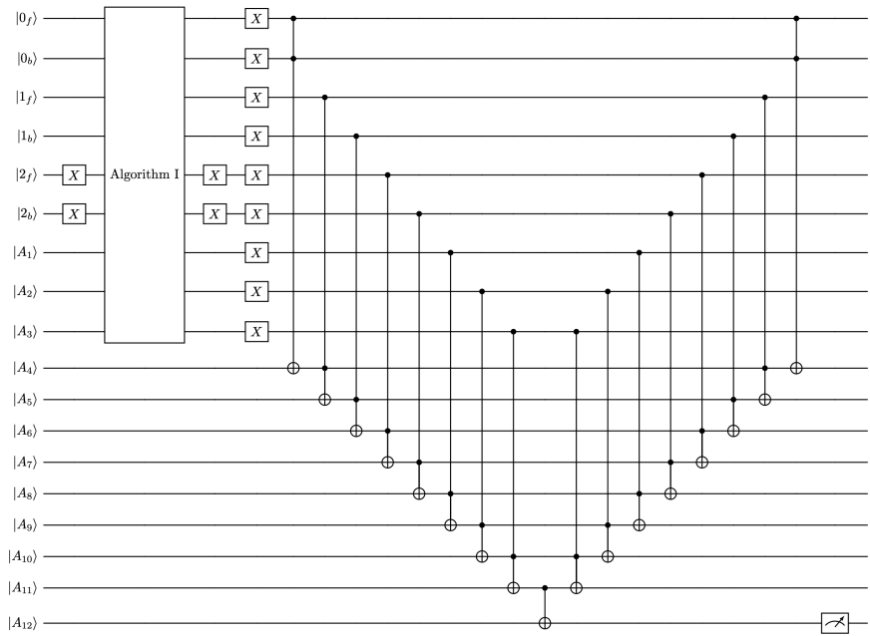

Figure 7. One-qubit measurement scheme for 2 time steps evolution for population $|1\rangle$.

The pros and cons for each of the algorithms are transparent. Algorithm I uses less number of qubibts whereas Algorithm II does less measurement. Indeed, for Algorithm II, only one qubit is measured regardless of the propagation time. It has the benefit of reducing the Monte Carlo sampling error from multiple qubit measurement, however, at the sacrifice of demanding more qubits. The simulation results from both of the algorithms are presented in the Section III.

## D. Complexity analysis
### (a) Algorithm I

For an $n$ level system, $log_2 n$ number of qubits is needed. Because the path integral has forward and backward states, it doubles the amount of qubits, i.e., $2log_2 n$. With the propagation time $N\Delta t$, there are $N + 1$ such pairs of forward backward states. Therefore, the number of qubits needed is $2(N + 1)log_2 n$. In addition, the number of ancila qubits required for dilation is equal to the number of compact two-timepoint coupling operators. The number of compact two-timepoint coupling operators is equal to the number of all possible pairwise couplings between timepoints, which is $N + (N - 1) + \cdots + 1 = (N + 1)N/2$. Putting these together, the total number of qubits needed for Algorithm I for an $n$ level system propapationg up to $N\Delta t$ is

$$2(N + 1)log_2 n + \frac{(N + 1)N}{2} \qquad (3.20)$$

As a matter of fact, in many condensed phase environment, the system has only finite memory. Suppose the memory length is $L\Delta t$. It means the timepoint coupling can span at most $L\Delta t$, where $L \leq N$. In this case, the number of compact two-timepoint coupling operators is $N + (N - 1) + \cdots + (N - L + 1) = (2N - L + 1)L/2$. Therefore, the total number of qubits needed for Algorithm I for an $n$ level system propapationg up to $N\Delta t$ with memory length $L\Delta t$ is

$$2(N + 1)log_2 n + \frac{(2N - L + 1)L}{2} \qquad (3.21)$$

The circuit depth soley arises from the implementation of the diagonal matrix of the compact operator $I^C$, which can be construted with the Walsh operators discussed in Section B. The total number of native gates for an $M$-qubit system is $2^{M+1} - 3$.[42] These native gates only contain one-quibt $R_z$ rotational gate and the two-qubit CNOT gate. Each $I^C$ operator couples two timepoints and requires $M = 4log_2 n + 1$ number of qubits. The $+1$ term accounts for the dilation. For propagation up to $N\Delta t$, there are $(N + 1)N/2$ such $I^C$ operators. Therefore, the total number of native gates are $(2^{4log_2 n+1 +1} - 3) \times (N + 1)N/2$. The additional Hadamard gates and the X gates are of constant depth (a small number) and does not affect the asymptotic behavior of the scaling. Therefore, the circuit depth for Algorithm I for an $n$ level system propapationg up to $N\Delta t$ is

$$(4n^4 - 3) \times \frac{(N + 1)N}{2} \qquad (3.22)$$

If the system has memory length of $L\Delta t$, the circuit depth becomes

$$(4n^4 - 3) \times \frac{(2N - L + 1)L}{2} \qquad (3.23)$$

In conclusion, for an $n$ level system propapationg up to $N\Delta t$, Algorithm I demands $O(log_2 n)$ and $O(N^2)$ number of qubits. If the memory length $L < N$, $O(N^2)$ is reduced to $O(N)$. The circuit depth scales as $O(n^4 N^2)$, and if the memory length $L < N$, the scale reduces to $O(n^4 N)$.

### (b) Algorithm II

For Algorithm II, because of the Toffoli gates, the number of qubits doubles exactly. With respect to the circuit depth, each Toffili gate can be decomposed into a constant number of native single and CNOT gates. The number of Toffili gate is equal to $2\times$(number of qubits in Algorithm One $-$ 2), resulting in $O[4(N+1)log_2 n + (N+1)N]$, or $O[4(N+1)log_2 n + (2N-L+1)L]$ if the $L < N$. Therefore, Algorithm II does not change the asymptotic behavior of Algorithm I with respect to both the number of qubits and the circuit depth.

In summary, both the algorithms we have prescribed scale polynomially with the number of qubits and the number of native gates.

## IV. Results and Discussions

In this section, we present our simulation results. We use the spin-boson model to demonstrate the quantum algorithm thus proposed. Specifically, the system Hamiltonian describes a symmetric two-level system with a non-zero off-diagonal coupling,

$$H_s = -\hbar\Omega(|s_1\rangle\langle s_2| + |s_2\rangle\langle s_1|) \tag{4.1}$$

where $|s_1\rangle$ and $|s_2\rangle$ are localized states that are eigenstates of the position operator $\hat{s}$ as well as the Pauli operator $\sigma_z$,

$$\hat{s}|s_i\rangle = s_i|s_i\rangle = \sigma_z|s_i\rangle \tag{4.2}$$

These states $|s_i\rangle$ are often called discrete value representation states.[45–47]

The system-bath coupling is given by

$$H_{sb} = \sum_j \left[\frac{p_j^2}{2m_j} + \frac{1}{2}m_j\omega_j^2\left(Q_j - \frac{c_j\sigma_z}{m_j\omega_j^2}\right)^2\right] \tag{4.3}$$

We choose the spectral density to have the Ohmic form,

$$J(\omega) = \frac{\pi}{2}\hbar\xi\omega e^{-\omega/\omega_c} \tag{4.4}$$

which gives a continuous version of equation (2.4).

We present results for two sets of parameters, as shown in Figure 8 and 9. The exact benchmark is obtained from the numerically exact tensor-network path integral method.[48] The ideal simulation is obtained by simulating the circuits using the *AerSimulator* with the native gates of *ibm_kyiv*. The noisy simulation is obtained from the real-time noise profile of *ibm_kyiv*. The dot in the graph is the average from 100 runs, and the error bar is the standard deviation from those 100 runs. The details of the number of shots and runs are given in Table I. The ideal simulator results for both the Algorithm I and II are in excellent agreement with the exact calculations. The presence of error bars on the ideal simulation is due to the shot noise. The noisy simulator results are of acceptable accuracy when considering the level of noise on NISQ processors. The noisy result of Algorithm II is slightly worse than that of Algorithm I due to the higher number of qubits involved. The *ibm_kyiv* machine and *AerSimulator* have a coupling map of 33 qubits, which limits our ability to propagate further in time. In the case of Algorithm I, we can propagate up to 5 time steps, and for

Algorithm 2, we can only demonstrate up to 3 time steps. The noisy simulations can not be carried out to the same extent as the ideal ones due to the large requirement of RAM on the classical computer.

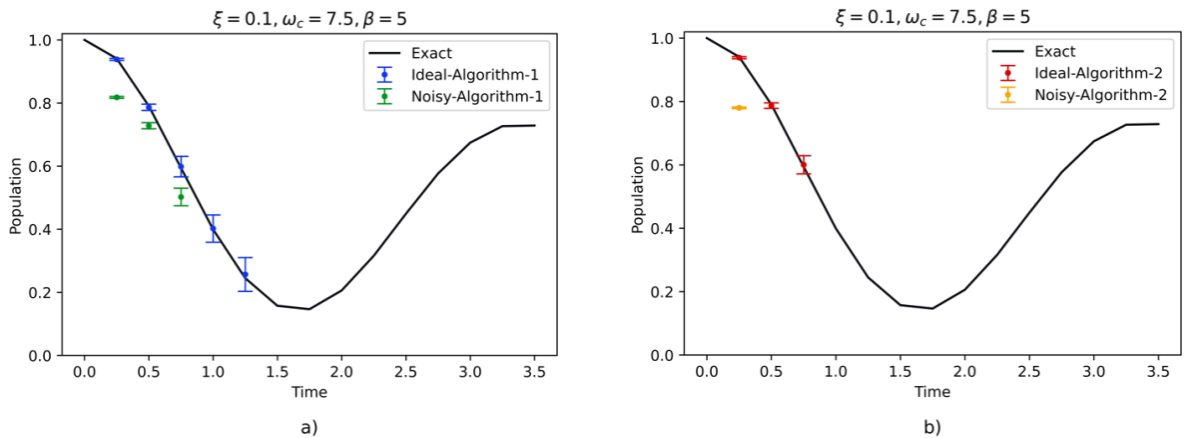

Figure 8. Population dynamics of the spin-boson model with parameters $\Omega = 1$, $\xi = 0.1$, $\omega_c = 7.5$, $\beta = 5$. Figure (a) is from Algorithm I and Figure (b) is from Algorihtm II. The ideal simulation is obtained by simulating the circuits using the *AerSimulator* and the noisy simulation is obtained from the real-time noise profile of *ibm_kyiv*.

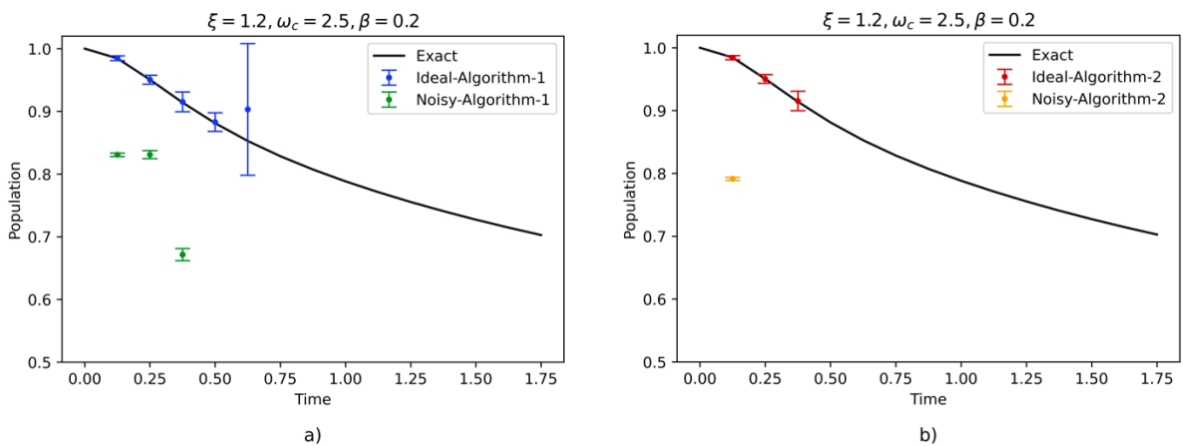

Figure 9. Population dynamics of the spin-boson model with parameters $\Omega = 1$, $\xi = 1.2$, $\omega_c = 2.5$, $\beta = 0.2$. Figure (a) is from Algorithm I and Figure (b) is from Algorihtm II. The ideal simulation is obtained by simulating the circuits using the *AerSimulator* and the noisy simulation is obtained from the real-time noise profile of *ibm_kyiv*.

| | Algorithm I | | | | Algorithm II | | | |
|---|---|---|---|---|---|---|---|---|
| | $\xi = 0.1, \omega_c = 7.5, \beta = 5$ | | $\xi = 1.2, \omega_c = 2.5, \beta = 0.2$ | | $\xi = 0.1, \omega_c = 7.5, \beta = 5$ | | $\xi = 1.2, \omega_c = 2.5, \beta = 0.2$ | |
| NUMBER OF TIME STEPS | Ideal Simulator | Noisy Simulator | Ideal Simulator | Noisy Simulator | Ideal Simulator | Noisy Simulator | Ideal Simulator | Noisy Simulator |
| | ×1000 shots | ×1000 shots | ×1000 shots | ×1000 shots | ×1000 shots | ×1000 shots | ×1000 shots | ×1000 shots |
| 1 | 20 | 20 | 20 | 20 | 20 | 20 | 20 | 20 |
| 2 | 30 | 30 | 75 | 75 | 30 | | 75 | |
| 3 | 40 | 40 | 300 | 300 | 40 | | 300 | |
| 4 | 250 | | 5000 | | | | | |
| 5 | 5000 | | 5000 | | | | | |

Table 1. Number of shots used in each time step.

According to the central limit theorem, for a fixd number of shots, with enough number of runs, the average of the ideal simulation will converge to the exact result. This is shown in Figure 10.

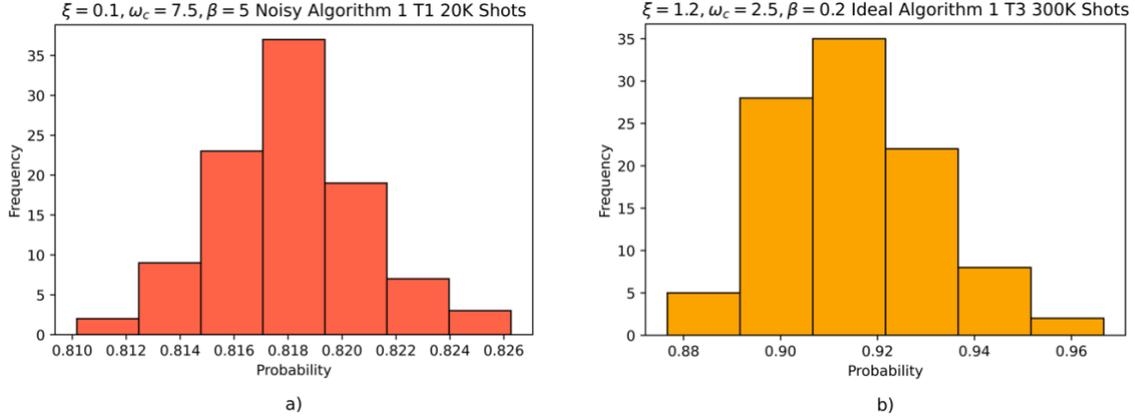

Figure 10. Representative Gaussian distribution. Figure (a) is from the noisy simulation of Algorith I at $T = \Delta t$ with parameters $\Omega = 1$, $\xi = 0.1$, $\omega_c = 7.5$, $\beta = 5$, 20K shots and 100 runs. Figure (b) is from the ideal simulation of Algorith I at $T = 3\Delta t$ with parameters $\Omega = 1$, $\xi = 1.2$, $\omega_c = 2.5$, $\beta = 0.2$, 300K shots and 100 runs.

We have also compared the standard deviation of Algorithm I and II, shown in Figure 11. As can be seen, Algorithm II does consistently better than Algorithm I. The reason is that Algorithm I presents its ending state as $\sum_i A_i |00 \cdots 00\rangle + \alpha |00 \cdots 01\rangle + \beta |00 \cdots 10\rangle + \cdots$, whereas Algorithm II has its ending state as $\sum_i A_i |0\rangle + (\alpha + \beta + \cdots)|1\rangle$. Each of the complex amplitudes in $(\alpha + \beta + \cdots)$ can interfere and the amplitude spared, $(\alpha + \beta + \cdots)^2$, gives a smaller value than $\alpha^2 + \beta^2 + \cdots$ in Algorithm I. Therefore, the measurement result of $|0\rangle$ in Algorithm II becomes more significant than that of $|00 \cdots 00\rangle$ in Algorithm I. It is evident from Table I that the number of shots needed increases as time progresses, as more terms populating into the term $(\alpha + \beta + \cdots)$ making the amplitude $\sum_i A_i$ relativaley smaller. Oblivious amplitude amplification[49] can be used to boost up the amplitude without changing the time and space complexty.

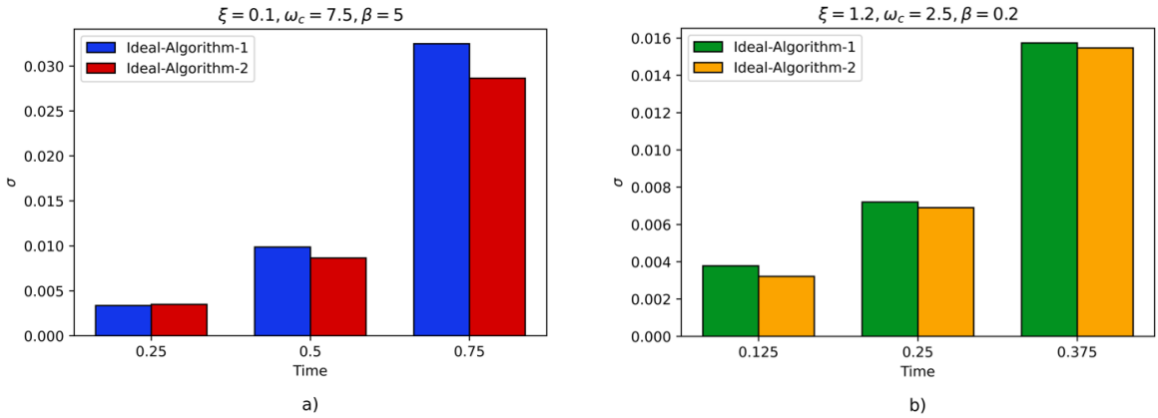

Figure 11. Standard deviation comparison for both algorithms with 100 runs. Figure (a) with paremeneters $\Omega = 1$, $\xi = 0.1$, $\omega_c = 7.5$, $\beta = 5$ and Figure (b) with parameters $\Omega = 1$, $\xi = 1.2$, $\omega_c = 2.5$, $\beta = 0.2$.

It is promising that with the rapid technological improvements[50–52] and the emerging new platforms,[53,54] quantum computers in the foreseeable future should be able to simulate non-Markovian quantum dynamics that are prohibitively expensive on classical computers.

## V. Conclusion

In this work, we have developed an efficient quantum algorithm with polynomial time and space complexity for simulating non-Markovian quantum dynamics that overcomes the exponential scaling in classical computing. Its advantage is particularly significant in the regime of strong non-Markovianity (i.e., high temporal entanglement) where tensor network methods fail. The algorithm is based on Feynman path integral's tensor product structure, and each operator couples two timepoints within the memory span. This unique feature allows a very compact circuit scheme with only a few qubits encoding each matrix. In addition, with the Wash operator representing the unitary diagonal, the circuit is both shallow and exact. We proposed two algorithms, one being qubit efficient, the other measurement effective. We tested the algorithms using the spin-boson model and the results from the ideal simulator match very well with the exact benchmark results. We are confident that the algorithm can be naturally adapted to simulate multi-level systems, and therefore demonstrates the quantum advantage for simulating non-Markovian quantum dynamics. These findings will prompt the quantum computer simulation of charge and exciton dynamics in the condensed phase environment, ranging from solution phase to biological systems.

**Acknowledgements**

This work is supported by the National Science Foundation (NSF) under the Award 2320328, and the George Mason University's startup fund. This work used the SDSC Expanse CPU of the Explore ACCESS through allocation CHE220009 and CHE240132 from the Advanced Cyberinfrastructure Coordination Ecosystem: Services & Support (ACCESS) program, which is supported by National Science Foundation grants 2138259, 2138286, 2138307, 2137603, and 2138296. We also acknowledge the use of IBM Quantum services for this work. The views expressed are those of the authors and do not reflect the official policy or position of IBM or the IBM Quantum team.

**Appendix**

**A.** Below $(a.1 - a.6)$ are the coefficients appeared in the influence functional in equation (3.4).

$$\alpha_{k'k} = \frac{2}{\pi}\int_{-\infty}^{\infty} d\omega \frac{J(\omega)}{\omega^2} \frac{\exp\left(\frac{\beta\hbar\omega}{2}\right)}{\sinh\left(\frac{\beta\hbar\omega}{2}\right)} \sin^2\left(\frac{\omega\Delta t}{2}\right) e^{-i\omega\Delta t(k-k')}, \qquad 0 < k < k' < N \qquad (a.1)$$

$$\alpha_{kk} = \frac{1}{2\pi}\int_{-\infty}^{\infty} d\omega \frac{J(\omega)}{\omega^2} \frac{\exp\left(\frac{\beta\hbar\omega}{2}\right)}{\sinh\left(\frac{\beta\hbar\omega}{2}\right)} \left(1 - e^{-i\omega\Delta t}\right), \qquad 0 < k < N \qquad (a.2)$$

$$\alpha_{N0} = \frac{2}{\pi}\int_{-\infty}^{\infty} d\omega \frac{J(\omega)}{\omega^2} \frac{\exp\left(\frac{\beta\hbar\omega}{2}\right)}{\sinh\left(\frac{\beta\hbar\omega}{2}\right)} \sin^2\left(\frac{\omega\Delta t}{4}\right) e^{-i\omega\left(t-\frac{\Delta t}{2}\right)}, \qquad (a.3)$$

$$\alpha_{00} = \alpha_{NN} = \frac{1}{2\pi}\int_{-\infty}^{\infty} d\omega \frac{J(\omega)}{\omega^2} \frac{\exp\left(\frac{\beta\hbar\omega}{2}\right)}{\sinh\left(\frac{\beta\hbar\omega}{2}\right)}\left(1 - e^{-\frac{i\omega\Delta t}{2}}\right), \qquad (a.4)$$

$$\alpha_{k0} = \frac{2}{\pi}\int_{-\infty}^{\infty} d\omega \frac{J(\omega)}{\omega^2} \frac{\exp\left(\frac{\beta\hbar\omega}{2}\right)}{\sinh\left(\frac{\beta\hbar\omega}{2}\right)} \sin\left(\frac{\omega\Delta t}{4}\right)\sin\left(\frac{\omega\Delta t}{2}\right)e^{-i\omega\left(k\Delta t - \frac{\Delta t}{4}\right)}, \qquad 0 < k < N \qquad (a.5)$$

$$\alpha_{Nk} = \frac{2}{\pi}\int_{-\infty}^{\infty} d\omega \frac{J(\omega)}{\omega^2} \frac{\exp\left(\frac{\beta\hbar\omega}{2}\right)}{\sinh\left(\frac{\beta\hbar\omega}{2}\right)} \sin\left(\frac{\omega\Delta t}{4}\right)\sin\left(\frac{\omega\Delta t}{2}\right)e^{-i\omega\left(t - k\Delta t - \frac{\Delta t}{4}\right)}, \qquad 0 < k < N \qquad (a.6)$$

The spectral density is extended to the negative frequencies defined as $J(-\omega) = -J(\omega)$ to avoid the singularity in the integration.

**B.** Below are the Walsh operator gate constructions for 3 and 5 qubit systems. The $G_i$ in the graph groups the binary strings of the qubit states into the subsets with most significant non-zero bit. [42]

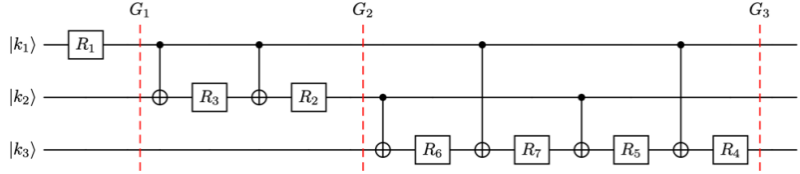

Figure B1. Optimal circuit for all 7 Wash operators on 3 qubits.

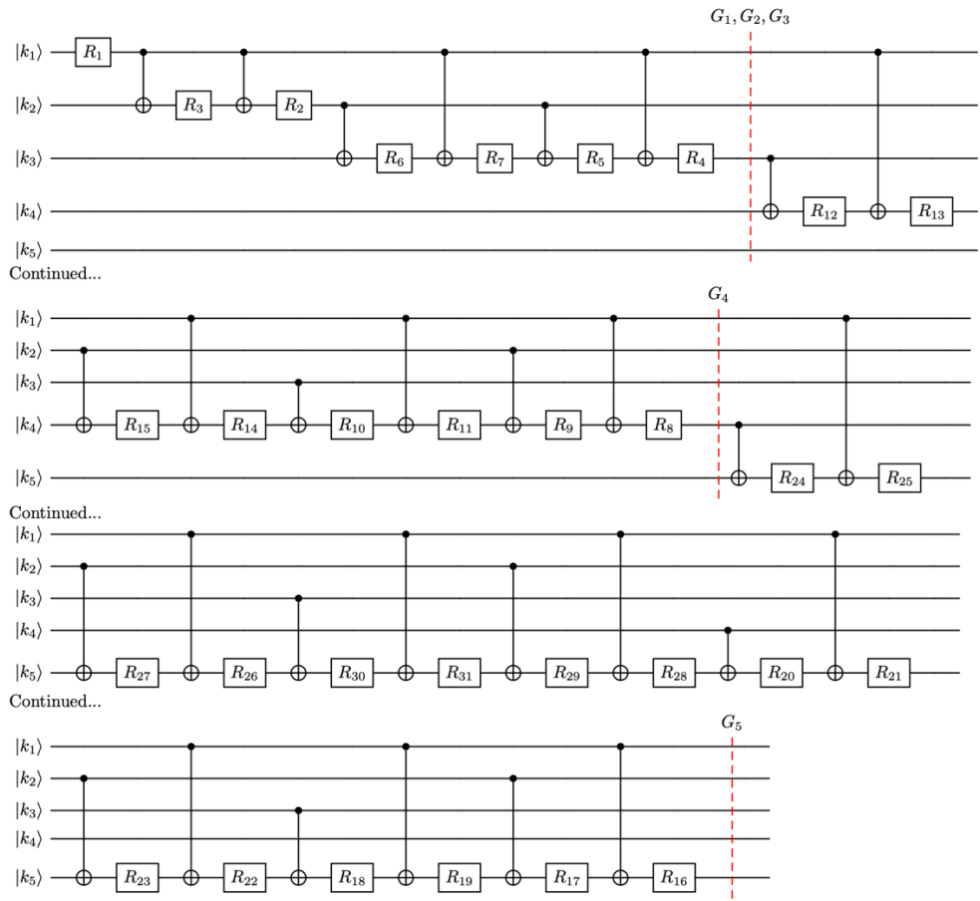

Figure B2. Optimal circuit for all 31 Walsh operators on 5 qubits.